\documentclass[12pt]{article}
\usepackage{blindtext}
\usepackage{multicol}
\usepackage{bbm}
\usepackage{graphicx}
\usepackage{lipsum}
\usepackage{microtype}
\usepackage{makecell}
\usepackage{url}
\usepackage{afterpage}
\usepackage{float}
\usepackage{titlesec}
\titlespacing\section{0pt}{5pt plus 1pt minus 2pt}{5pt plus 2pt minus 2pt} 
\titlespacing\subsection{0pt}{5pt plus 1pt minus 2pt}{5pt plus 2pt minus 2pt} %
\DisableLigatures[-]{}
\newcommand{\project}{FrameCorr}
\newenvironment{keywords}{%
  \textbf{Keywords:}%
}

\usepackage[a4paper, total={7in, 10in}]{geometry}

\makeatletter
\renewcommand{\@maketitle}{%
  \newpage
  \null
  \vskip 2em%
  \begin{center}%
  \let \footnote \thanks
    {\LARGE \@title \par}%
    \vskip 1.5em%
    {\large
      \lineskip .5em%
      \begin{tabular}[t]{c}%
        \@author
      \end{tabular}\par}%
    \vskip 1em%
  \end{center}%
  \par
  \vskip 1.5em}
\makeatother

\title{FrameCorr: Adaptive, Autoencoder-based Neural Compression for Video Reconstruction in Resource and Timing Constrained Network Settings} 
\author{John Li, Shehab Sarar Ahmed, Deepak Nair}

\begin{document}
\maketitle

\begin{multicols*}{2}
\begin{abstract}
Despite the growing adoption of video processing via Internet of Things (IoT) devices due to their cost-effectiveness, transmitting captured data to nearby servers poses challenges due to varying timing constraints and scarcity of network bandwidth. Existing video compression methods face difficulties in recovering compressed data when incomplete data is provided. Here, we introduce FrameCorr, a deep-learning based solution that utilizes previously received data to predict the missing segments of a frame, enabling the reconstruction of a frame from partially received data.
\end{abstract}
\begin{keywords}
Video Transmission,
Progressive Compression,
IoT
\end{keywords}
\section{Introduction}
Video-enabled IoT devices provide a comprehensive view of the environment by capturing visual data alongside traditional sensor data, facilitating real-time monitoring and decision-making across various domains. However, due to resource constraints, these devices often rely on edge servers for video processing. This reliance introduces timing constraints that may require interrupting frame transmission to transition to the next frame. Consequently, edge servers frequently face the challenge of reconstructing video frames from incomplete data. Thus, there is a pressing need for an efficient method on the server side to effectively handle missing data in video frames.

One way of handling this challenge is to encode each frame of the video into a compressed form before transmission.
There has been numerous compression techniques, both classical (Huffman Coding \cite{huffman1952method}, JPEG \cite{wallace1991jpeg}, MPEG \cite{le1991mpeg}, H.261 \cite{turletti1993h}, H.263 \cite{rijkse1996h}, H.264 \cite{wiegand2003overview}, HEVC \cite{sullivan2012overview}) and neural network based ones using multilayer perceptrons (MLPs) \cite{chua1988neural,sicuranza1990artificial}, Convolutional Neural Networks (CNNs) \cite{balle2016end, ahanonu2018lossless} and AutoEncoders \cite{theis2017lossy}, which reduce overall frame size.
However, none of these methods is explicitly designed to manage decompression with partially received data, which often becomes the only recourse when the sender is unable to transmit complete data due to shortages in time.

In one very recent method, Progressive Neural Compression (PNC)  \cite{wang2023progressive}, the authors propose a progressive encoding of images that can tolerate missing data.
However, it only applies to static images and relies on zero-filling to address missing data, a method that may result in suboptimal performance for videos due to not leveraging the inter-frame correlation between consecutive frames.

This paper presents \textit{\project}, a deep-learning framework designed to exploit inter-frame correlation for efficiently reconstructing missing data within a frame.
Additionally, we implemented our own version of adaptive bitrate (ABR) video delivery on top of AVC to juxtapose its performance with that of \project, aiming to highlight differences in their methodologies. Unlike \project, which involves partitioning and extracting image frames and features from the videos, ABR solely adheres to the same video format without such extractions. We observed that AVC, when paired with ABR, outperforms deep learning-based methods in terms of both throughput and accuracy. Nonetheless, traditional algorithms like AVC exhibit limitations when confronted with incomplete data, rendering them unsuitable for tasks with strict timing requirements.

\section{Related Work}
\textbf{Conventional Image and Video Compression}. Traditional methods of compression are primarily designed to reduce data volume necessary for image restoration. There are lossy compression techniques such as the coefficient quantization step in JPEG's 3-stage compression algorithm and lossless methods like discrete cosine transform (DCT) transformation in the same JPEG compression process \cite{WinNT}. Furthermore, JPEG has a progressive mode, allowing images to be compressed and reconstructed more flexibly based on an arbitrary amount of encoded data, where more data produces a more accurate reconstruction \cite{hofer2023progressive}.

However, such compression algorithms are only applicable to static images. Video content is addressed through other compression protocols. Low latency video coding and compression is especially relevant in IoT/Edge computing systems due to the real time deadlines that these applications generally operate under. One such video compression method is the H.264 \cite{nemcic2007comparison, wiegand2003overview} codec, also referred to as AVC, a version of the MPEG standard that incorporates block-based motion compensation strategies and exploits spatial and temporal repetitions. H.264 uses a combination of I-frames (Intra frames: these are complete frames that contain full image information) and P-frames (Predictive frames: these encode the differences or changes between themselves and previous reference frames) for video compression. I-frames are periodically inserted in the video stream or at scene changes to provide starting points for decoding. P-frames exploit spatial and temporal redundancies by ``predicting" image content based on previously decoded frames.

\textbf{Adaptive Video Compression}
Adaptive video compression improves upon traditional methods by adjusting compression levels based on network conditions before transmission. A prominent example is adaptive bitrate (ABR) streaming, where video streams are compressed at various bitrates. Then, depending on the current network conditions, the appropriate bitrate-encoded videos are transmitted. Specific ABR protocols include HTTP Live Streaming (HLS) and Dynamic Adaptive Streaming over HTTP (DASH) \cite{HLSandDash}.

\textbf{Neural Compression and Inference}.
The advent of deep learning has ushered in innovative mechanisms for image compression, such as autoencoders (AEs) \cite{theis2017lossy}, nested quantization, latent ordering \cite{lu2021progressive}, and recurrent neural networks (RNNs) \cite{toderici2015variable}. These methods enable variable compression, allowing for incremental refinement of image quality. Despite their superior compression efficacy, these neural models often demand significant computational resources, sometimes requiring several hours on GPU clusters during training, making them impractical for resource-limited IoT and edge devices.

\textbf{Progressive Neural Compression.}
Neural compression techniques have evolved to include progressive compression strategies, essential for adapting to fluctuating bandwidth conditions common in wireless sensor networks and distributed IoT applications. One such recent approach, Starfish \cite{hu2020starfish}, introduces a method to enhance the resilience of neural compression to data transmission losses by adding random dropouts to its AE's bottleneck layer. Though Starfish does mitigate the impact of data loss, it lacks a mechanism for assessing and prioritizing the encoded features based on their importance for inference accuracy.

As aforementioned, PNC \cite{wang2023progressive} was developed to improve classification accuracy for images within edge offloading environments, particularly when faced with temporal and bandwidth limitations. Diverging from existing methodologies, PNC dynamically adjusts to changes in bandwidth, allowing for efficient image classification by the edge server. It does so by training a multi-objective rateless autoencoder, tailored for multiple compression rates. PNC also implements a stochastic taildrop algorithm during training to form a compression solution that creates features ordered by importance in the inference process.
However, PNC is designed to work with static images, hence does not leverage the inherent correlation between video frames in the process of filling up the missing data.

\section{System Model}
\begin{figure*}
\centering
    \includegraphics[width=.9\textwidth]{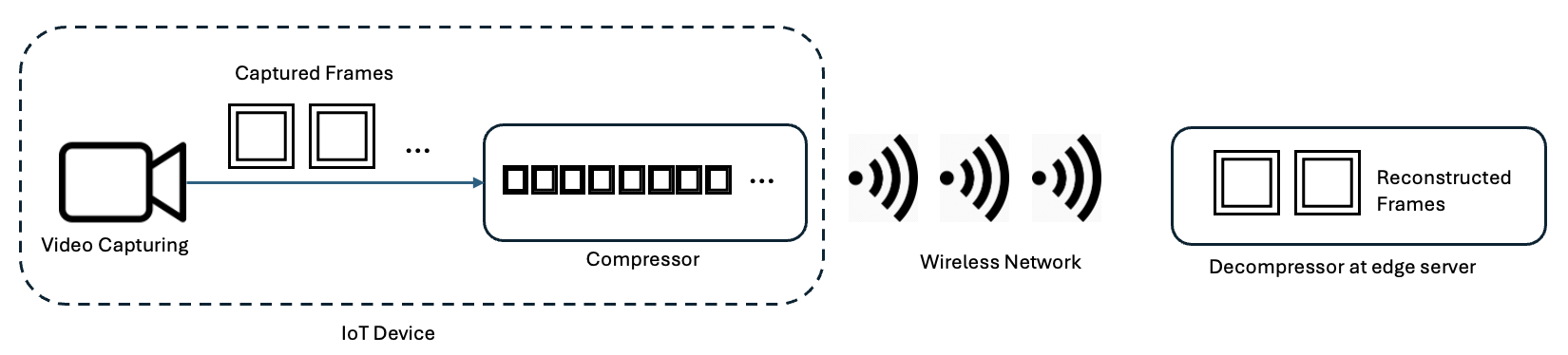}
 \caption{The system model we consider for video transmission entails the following process: initially, a video capturing device captures frames and subsequently compresses them. These compressed frames are then transmitted via the wireless network to the edge server. At the server, decoding is applied to reconstruct the frames as accurately as possible to the originals. 
 }
 \label{fig:arch}
\end{figure*}

Our system architecture, as depicted in Figure \ref{fig:arch}, includes a resource-constrained system (e.g., an IoT device or low-power virtual machine), responsible for transmitting compressed video data. This data is then sent over a wireless network to a central edge server. At the server, a decompression algorithm—such as PNC or \project—is utilized to reconstruct the received compressed data (usually in bytes) into their original video or image frame formats as accurately as possible.

An essential consideration in the transmission of compressed frames is adherence to strict deadlines at the sender's end. Due to potentially limited network bandwidth, the client device may still be sending a frame when the subsequent frame becomes ready for transmission. Consequently, we enforce a deadline for each frame's transmission, requiring the transmission of the next frame commence promptly, even if a portion of the current frame remains unsent. As a result, the receiving server must be able to robustly reconstruct frames based on partially received data, a key component of \emph{\project}. 

\section{Methods}
\subsection{Dataset}\label{subsec:dataset}
Our dataset was derived from the UCF Sports Action dataset \cite{rodriguez2008action, soomro2015action}, which originally comprises videos showcasing 10 distinct actions. From this dataset, we selected 8 of the 10 actions. Subsequently, we partitioned the videos from each action into training, validation, and test sets. The distribution of videos across each category is detailed in Table \ref{tab:dataset}.
\begin{table*}[t!]
    \centering
    \begin{tabular}{|l|p{2cm}|p{2cm}|p{2cm}|}
    \hline
    \textbf{Action}&\textbf{Train}&\textbf{Validation}&\textbf{Test}\\\hline
    Dive&4&1&2\\\hline
    Golf Front&4&2&2\\\hline
    Kick Front&6&2&2\\\hline
    Lift&3&1&2\\\hline
    Ride Horse&7&3&2\\\hline
    Run&7&3&3\\\hline
    Skate&7&3&2\\\hline
    Swing Bench&12&5&3\\\hline
    \end{tabular}
    \caption{Distribution of videos across the training, validation, and test sets for each of the 8 classes utilized in our experiments.}
    \label{tab:dataset}
\end{table*}
\subsection{AVC}
Initially, we selected the AVC/H.264 codec as our baseline for video compression due to its widespread use in numerous applications, minimal data loss, and ability to maintain the original video format without the need for additional frame conversion. However, it exhibits limited error-resilience, especially in network environments with unstable bandwidth or strict timing constraints. Furthermore, AVC mandates transmitting the entire encoded video over the network; dynamic transmission of partial video segments isn't possible. 

For our experiments, we utilized the FFmpeg library with .mp4 as the video format. 

\subsection{ABR}

To expand on the baseline H.264 method, we developed our own adaptive bitrate (ABR) video transmission implementation, which utilizes H.264 as the base encoding method and encodes content at various bitrates, specifically tweaking the control rate factor (CRF). In the FFmpeg library, a lower CRF (e.g. CRF=18) denotes a higher bitrate while a higher CRF (e.g. CRF=30) indicates a lower bitrate.

\subsection{PNC}
PNC, as described in \cite{wang2023progressive}, is a progressive encoding framework primarily designed for images. The PNC model undergoes a two-step training process. Initially, an autoencoder is trained to reconstruct images with high fidelity. Subsequently, the autoencoder is fine-tuned to optimize the accuracy of an image classifier using the reconstructed images. Throughout the training process, a stochastic tail-drop technique is employed to enhance the autoencoder's ability to reconstruct images from partially received data. In this technique, missing data is padded with zeros, ensuring that the decoder receives a fixed-size vector.

In our context, our focus is precisely on the reconstruction of video frames. Thus, we tailor the training of the autoencoder to prioritize minimizing the reconstruction error of these frames. To formalize this, let $x_i$ denote the $i^{th}$ captured frame, and $E(x_i)=c_i$ represent the encoded data of $x_i$, computed by the IoT device. These encoded representations, denoted as $c_i$, are transmitted over the network. However, due to timing constraints, the sender may opt to switch to encoding and sending the next frame, thereby interrupting the transmission of the current frame. Consequently, the received data for the $i^{th}$ frame is denoted as $\dot{c}_i$. Upon receiving $\dot{c}_i$, the receiver zero-pads it to match the dimension of $c_i$, resulting in $\hat{c}_i$. Subsequently, $\hat{c}_i$ is passed through the decoder to reconstruct the frame, denoted as $\hat{x}_i$. PNC is trained to minimize the mean squared error (MSE) between the original frame ($x_i$) and its reconstructed counterpart ($\hat{x}_i$).

In conclusion, although consecutive frames within a video usually exhibit correlation, PNC, being originally designed for images, does not leverage this correlation. Consequently, missing data is filled with zeros during reconstruction.
\begin{figure*}[htbp!]
\centering
    \includegraphics[width=.9\textwidth, height=0.4\textheight]{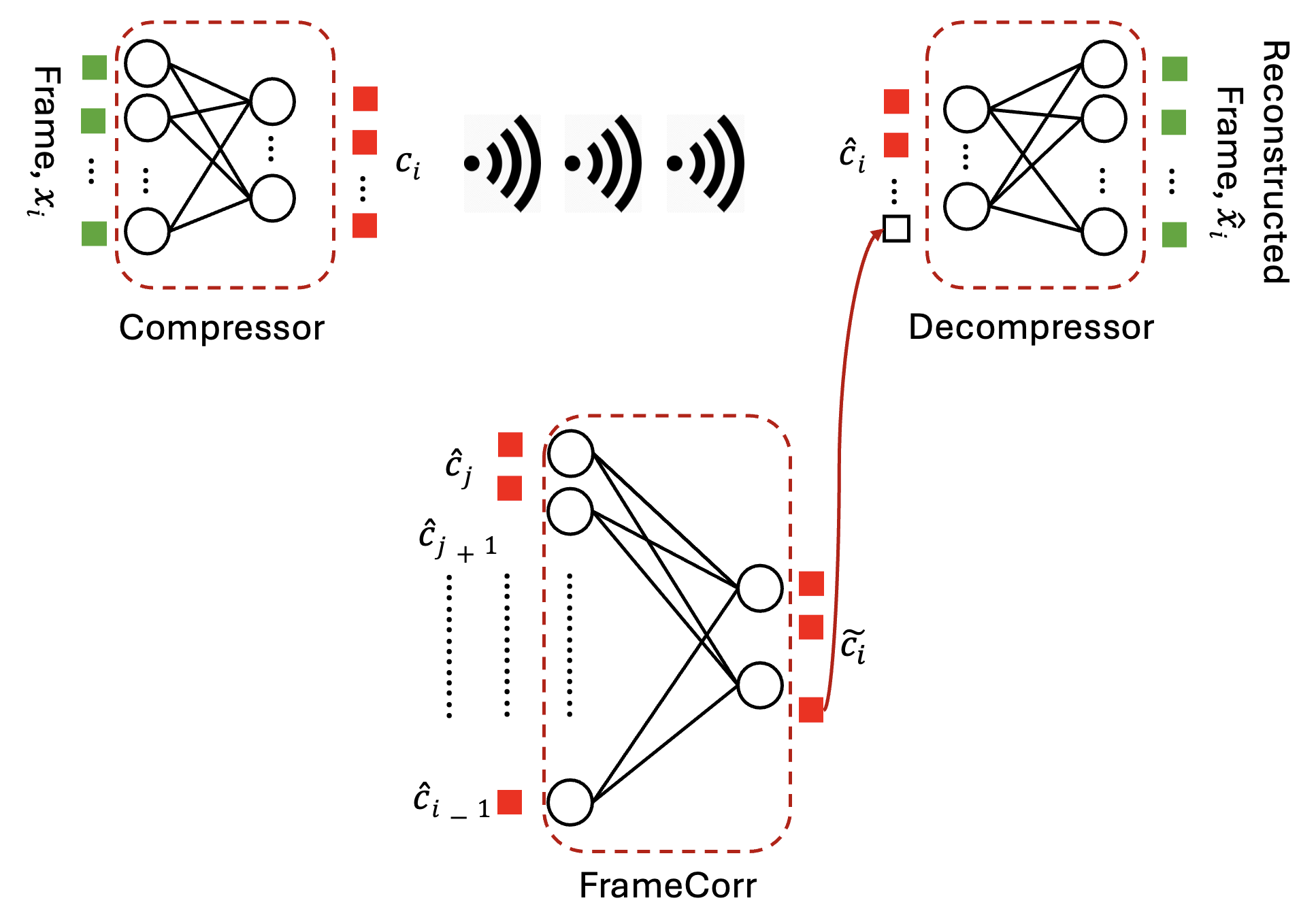}
 \caption{Every frame $x_i$ undergoes compression to yield $c_i$, which is then transmitted across the network. Upon reception by the server, PNC zero-pads the received data to align with the dimensions of $c_i$. On the other hand, we develop a distinct deep-learning model, referred to as \project, to predict the encoded details of the current frame, denoted as $\tilde{c}_i$, based on the encoded information of the preceding $K$ frames. The absent segments of $c_i$ are populated with the corresponding portions of $\tilde{c}_i$, denoted as $\hat{c}_i$.}
 \label{fig:framecorr}
\end{figure*}
\subsection{\project}
As previously mentioned, given the inherent correlation present in consecutive frames, our aim is to enhance PNC by leveraging this relationship to fill in missing data rather than simply padding it with zeros. 
This is where \project ~comes into play. 
The architecture of \project ~along with the PNC autoencoder is illustrated in Figure \ref{fig:framecorr}. The \project ~model takes as input the encoded information from the preceding $K$ frames ($\hat{c_j}$'s). It then predicts the encoded information for the current frame ($\tilde{c_i}$). To address the possibility of missing data, stochastic taildrop is applied to the $\hat{c_j}$'s in the training phase. When there is a missing segment in the received data, the missing parts are filled up to produce $\hat{c_i}$, using the predicted value from the output of \project's decoder component.




\section{Results}
\subsection{Experimental Setup}
Our experimental setup utilized two virtual machines (VMs) within a remote cluster farm, modeling the IoT device and edge server in our system. Each VM was provisioned with 2 CPU cores, 4GB RAM, and 100GB of storage, running on the Red Hat Enterprise Linux 8 (64-bit) operating system. To match the software environment of the original PNC paper \cite{wang2023progressive}, we configured the system with Python 3.8.7, TensorFlow 2.8 and other matching packages.

The VM emulating the IoT device ran the encoder, while the VM emulating the edge server handled the decoder and frame reconstruction. For networking, we employed a custom TCP connection initiated by the IoT device using the Python sockets library. All frames were iteratively passed through this connection. Each frame was compressed and individually chunked into packets, which were then combined and decompressed at the edge server.

To facilitate the reception of smaller frames without socket blocking, a 3-byte delimiter was added for identification. Additionally, a 3-byte ACK delimiter was used as an acknowledgment signal, allowing the sender and receiver to coordinate when data transmission is permitted.

Network conditions were varied using the Linux Traffic Control Toolkit, a command-line tool that simulates network behavior such as delays, packet loss, and bandwidth limitations. A shell script modified the system to pre-set network configurations modeling different network qualities before initiating video transfer. 

Specifically, we tested with a wide range of network conditions. However, for clarity and simplicity in our experiments, we categorized the network conditions into three levels: minimal, medium, and high congestion. The main adjustable parameters in the Linux Traffic Control Toolkit are the data rate (limiting the maximum bandwidth available), burst size (defining the initial amount of data that can be sent at higher speeds before throttling to the set data rate), and latency (the time a packet is held in the buffer before getting processed or dropped). For high network congestion, the data rate was capped at 1 megabit per second, the burst size was set at 32 kilobits, and the latency was set to 400 ms. For medium congestion, the rate was set at 10 megabits per second, with a burst size of 64 kilobits and a latency of 200 ms. Finally, the arguments for minimal congestion were a rate of 50 megabits per second, a burst size of 128 kilobits, and latency of 50 ms.

The general dataflow, including measurements for AVC, PNC, and FrameCorr, remained consistent across all experiments with slight modifications to accommodate specific algorithm requirements. For instance, zero-padding was implemented for missing data depending on the reconstruction algorithm used (required by PNC but not FrameCorr). Our code then measured mean squared error, network latency, bandwidth, and other system metrics.

We opted for a virtual machine (VM) environment in our experiments to gain precise control over network conditions. This approach provided flexibility to simulate various real-world IoT scenarios without the cost and complexity of physical hardware and wireless equipment. Future work involves migrating our testbed to real IoT devices operating on a wireless channel.

\subsection{Reconstruction with Complete Data}
We train both PNC and \project ~on the dataset mentioned in Section \ref{subsec:dataset} with the objective of minimizing the reconstruction error. The encoder output dimension is set to 10. During training, we utilize the validation dataset to compute the loss after each epoch. We checkpoint the model with the lowest validation loss observed thus far. The training was run for 15 epochs.

The number of bytes a video contains serves as an indirect metric for measuring overall throughput, where more bytes take longer to send. 
This correlates with network bandwidth as higher bandwidth allows for higher throughput and vice versa. 

\begin{figure*}
\centering
    \includegraphics[scale=0.9, height=0.42\textheight]{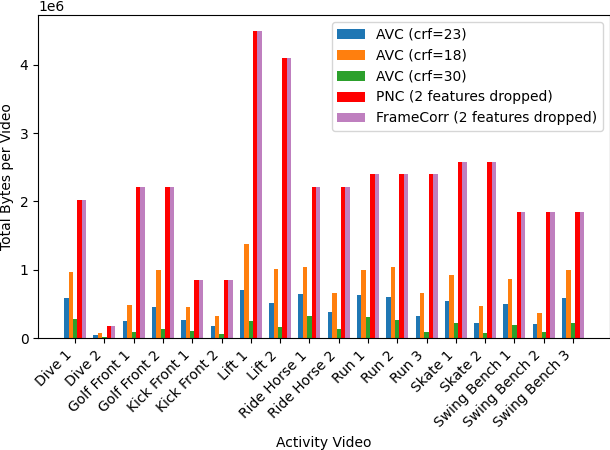}
 \caption{The average number of bytes per frame in the encoded information across the 18 videos in the test set}
\label{fig:trinket_plot_total_bytes_UPDATED}
\end{figure*}

We present the number of bytes of encoded information for the 18 videos in our test set for PNC, \project~, and AVC in Figure \ref{fig:trinket_plot_total_bytes_UPDATED}. It is notable that AVC consistently requires fewer bytes to encode compared to PNC or \project.
Additionally, it is noteworthy that the number of bytes of encoded information for PNC remains the same as \project ~across videos. This consistency is attributed to their fixed encoder output dimensions of 10.

The Mean Squared Error (MSE) is utilized as the metric to quantify the difference between a reconstructed frame and its original counterpart. Prior to passing through the encoding process, each pixel value is normalized to the range $[0,1]$.

We compute the MSE of each video by first summing the squared differences of the pixel values between the original and reconstructed frames and then averaging those differences with respect to the number of frames embedded in each video. The MSE achieved by PNC, FrameCorr (no feature drops), and AVC are reported in Table \ref{tab:mse}. It turns out the MSE achieved by AVC is consistently lower than those achieved by PNC and \project, showing the success of traditional compression methods when no data is dropped.

Lastly, we recorded the total time it took for each method to process the entire dataset of videos or video frames as per Table \ref{tab:latency}. It's vital to note that both PNC and \project ~took significantly longer (overall more than 3x greater than that of the highest AVC bitrate encoding whose CRF = 18) to process all the video frames. This is most likely due to the fact that 1) a sequence of extracted image frames consists of more bytes than the actual video themselves due to the innate compression techniques AVC embeds into the .mp4 videos and 2) the extra overhead associated with acknowledging packet transfer for not only every frame but also every feature. We also assumed that no timing constraint would be enforced, which also implies no feature vectors are dropped for PNC and \project. 

\begin{table*}[t!]
    \centering
    \begin{tabular}{|l|p{2.5cm}|p{2.5cm}|p{2.5cm}|p{2cm}|}
    \hline
    \textbf{\makecell{Video}} & \textbf{AVC H.264 (CRF=18)} & \textbf{AVC H.264 (CRF=23)} &  \textbf{AVC H.264 (CRF=30)}& \textbf{PNC and FrameCorr}\\\hline
    Dive 1 & 81.57 & 141.03  & 290.13 & 540.49\\\hline
    Dive 2 & 51.35 & 88.17  & 197.51&404.63\\\hline
    Golf Front 1 & 42.65 & 73.86  & 149.67 & 669.14\\\hline
    Golf Front 2 & 22.03 & 34.34  & 62.45 & 172.42\\\hline
    Kick Front 1 & 115.12 & 188.31  & 363.39 & 1,742.93\\\hline
    Kick Front 2 & 88.50 & 152.27  & 292.84 & 706.31\\\hline
    Lift 1 & 103.29 & 177.67  & 339.13& 976.91\\\hline
    Lift 2 & 84.14 & 145.45 & 284.85& 623.21\\\hline
    Ride Horse 1 & 76.39 & 128.71 & 274.39 & 468.35\\\hline
    Ride Horse 2 & 67.69 & 114.21 & 236.23 & 326.75\\\hline
    Run 1 & 43.51 & 73.61  & 155.14 & 154.96\\\hline
    Run 2 & 42.60 & 69.10 & 134.21 & 324.47\\\hline
    Run 3 & 84.33 & 140.94 & 272.07 & 590.45\\\hline
    Skate 1 & 75.84 & 135.38 & 289.49 & 739.66\\\hline
    Skate 2 & 41.01 & 67.00 & 126.13 & 254.72\\\hline
    Swing Bench 1 & 178.85 & 310.62 & 629.00 & 1,172.00\\\hline
    Swing Bench 2 & 51.71 & 89.84 & 173.88 & 241.40 \\\hline
    Swing Bench 3 & 215.73 & 370.18 & 758.60 & 1,339.71\\\hline
    \end{tabular}
    \caption{The respective mean squared errors (MSEs) attained by PNC (with 0 features dropped) and AVC (at various control rate factors (CRF) to accommodate adaptive bitrate streaming) without any loss of encoded data for videos in the test dataset.}
    \label{tab:mse}
\end{table*}

\begin{table*}[t!]
\centering
\begin{tabular}{|c|c|c|c|c|c|}
\hline
{ } & {Low Network Congestion} & {Medium Network Congestion} & {High Network Congestion} \\
\hline
{AVC (CRF=30)} & {0.799749 s} & {5.171132 s} & {24.79438 s} \\
\hline
{AVC (CRF=23)} & {2.485729 s} & {13.349195 s} & {61.154762 s} \\
\hline
{AVC (CRF=18)} & {7.049102 s} & {32.112081 s} & {102.704191 s} \\
\hline
{PNC} & {21.049102 s} & {89.112081 s} & {322.79438 s} \\
\hline
{FrameCorr} & {22.157273 s} & {91.918901 s} & {329.768233 s} \\
\hline
\end{tabular}
\caption{Latency or total time elapsed for each method to transmit all videos under various network conditions with no deadline/timing constraint enforced. This also means no features drop recorded for PNC and FrameCorr}
\label{tab:latency}
\end{table*}
\subsection{Reconstruction with Partial Data}

We quantified the percentage of video successfully transmitted for a single video clip under fluctuating network conditions and a set timing constraint (deadline), as documented in Table \ref{tab:percentage_of_vids_sent}. Notably, transmission success was binary: either 0\% or 100\% of the video was successfully transferred for AVC, reflecting the integral encoding structure of .mp4 videos. Attempting to send a ``partial" video would simply break its inherent structure and corrupt its data. In contrast, for PNC and FrameCorr, most if not all video frames were  transmitted. However, some features were omitted. Specifically, under minimal congestion, no features were dropped; under medium congestion, an average of 1-2 features per frame were omitted; and under high congestion, an average 3-4 features per frame were dropped.

AVC is unable to reconstruct videos with partial information since the codec requires complete data for the video to be properly decoded. As for PNC and \project, approximately 1-4 features (out of the 10 features) were dropped; we tasked the decoder with reconstructing the frame using the remaining features respectively and reported the MSE thereafter.
The MSE achieved by PNC and \project ~for these scenarios is presented in Tables \ref{tab:pnc_data_loss} and \ref{tab:FrameCorr_data_loss}. We report the results for \project ~trained with $K=1$, which gives the best MSE values among values of $K$ from 1 to 4. 


Unexpectedly, PNC surpasses \project ~in nearly all video instances, indicating that simply zero-padding the absent segments performs admirably across most scenarios. Moreover, the MSE values rise with the escalation of dropped features, as anticipated. However, the marginal increase observed in both PNC and \project ~demonstrates the resilience of these approaches when confronted with missing data events.

\begin{table*}[t!]
    \centering
    \begin{tabular}{|l|p{2.5cm}|p{2.5cm}|p{2.5cm}|p{2.5cm}|}
    \hline
    \textbf{Video} & \textbf{Drop 1 Feature} & \textbf{Drop 2 Features} & \textbf{Drop 3 Features} & \textbf{Drop 4 Features} \\\hline
    Dive 1 & 540.49 & 540.62256 & 540.5136 & 551.82806 \\\hline
    Dive 2 & 404.63083 & 404.71814 & 406.4142 & 421.58194 \\\hline
    Golf Front 1 & 669.137 & 669.15564 & 686.2272 & 696.08215 \\\hline
    Golf Front 2 & 172.42131 & 172.48883 & 172.52687 & 170.17465 \\\hline
    Kick Front 1 & 1742.931 & 1742.9437 & 1744.0746 & 1748.9518 \\\hline
    Kick Front 2 & 706.3137 & 706.42206 & 709.35944 & 710.65814 \\\hline
    Lift 1 & 976.9144 & 977.43243 & 972.99884 & 986.4046 \\\hline
    Lift 2 & 623.2105 & 623.6637 & 624.8001 & 637.1971 \\\hline
    Ride Horse 1 & 468.34634 & 468.9991 & 469.5005 & 475.9657 \\\hline
    Ride Horse 2 & 326.7453 & 326.83774 & 325.85703 & 335.35858 \\\hline
    Run 1 & 154.96104 & 154.97946 & 154.65768 & 158.85179 \\\hline
    Run 2 & 324.46808 & 325.1715 & 325.7509 & 327.75192 \\\hline
    Run 3 & 590.4475 & 593.3605 & 594.0178 & 616.4274 \\\hline
    Skate 1 & 739.6635 & 739.69714 & 741.7765 & 759.1474 \\\hline
    Skate 2 & 254.71878 & 254.73999 & 254.54688 & 263.20135 \\\hline
    Swing Bench 1 & 1172.0049 & 1172.0398 & 1174.114 & 1195.6516 \\\hline
    Swing Bench 2 & 241.39984 & 241.71928 & 243.60643 & 254.69327 \\\hline
    Swing Bench 3 & 1339.7064 & 1339.712 & 1342.9078 & 1371.6229 \\\hline
    \end{tabular}
    \caption{MSEs calculated for PNC, specifically the impact of dropping 1-4 features.}
    \label{tab:pnc_data_loss}
\end{table*}

\begin{table*}[t!]
    \centering
    \begin{tabular}{|l|p{2.5cm}|p{2.5cm}|p{2.5cm}|p{2.5cm}|}
    \hline
    \textbf{Video} & \textbf{Drop 1 Feature} & \textbf{Drop 2 Features} & \textbf{Drop 3 Features} & \textbf{Drop 4 Features} \\\hline
    Dive 1 & 626.26526 & 626.3737 & 629.2918 & 643.72046 \\\hline
    Dive 2 & 469.72577 & 469.86914 & 472.60123 & 490.6584 \\\hline
    Golf Front 1 & 819.14044 & 816.47186 & 820.9554 & 827.6751 \\\hline
    Golf Front 2 & 225.36108 & 223.88918 & 216.25337 & 222.65997 \\\hline
    Kick Front 1 & 1879.0557 & 1871.3534 & 1818.9274 & 1817.7074 \\\hline
    Kick Front 2 & 711.34705 & 709.91046 & 709.93243 & 723.4379 \\\hline
    Lift 1 & 827.77344 & 828.2437 & 828.3225 & 843.4022 \\\hline
    Lift 2 & 651.5644 & 653.4929 & 653.7797 & 665.87317 \\\hline
    Ride Horse 1 & 528.61163 & 529.6719 & 529.0928 & 542.60175 \\\hline
    Ride Horse 2 & 411.70258 & 413.73743 & 413.02255 & 419.88058 \\\hline
    Run 1 & 220.38101 & 220.12929 & 223.05417 & 227.29356 \\\hline
    Run 2 & 235.76015 & 234.90808 & 232.85027 & 241.7574 \\\hline
    Run 3 & 646.8063 & 652.8583 & 655.2728 & 665.3907 \\\hline
    Skate 1 & 881.4789 & 881.27893 & 885.2695 & 900.7622 \\\hline
    Skate 2 & 316.9019 & 316.6824 & 318.6439 & 326.03064 \\\hline
    Swing Bench 1 & 1268.3307 & 1268.1631 & 1271.7761 & 1291.3135 \\\hline
    Swing Bench 2 & 271.56995 & 271.67102 & 275.35602 & 284.75873 \\\hline
    Swing Bench 3 & 1402.5295 & 1402.2457 & 1406.4795 & 1430.8453 \\\hline
    \end{tabular}
    \caption{MSEs calculated by the {\project}, specifically the impact of dropping 1-4 features.}
    \label{tab:FrameCorr_data_loss}
\end{table*}

\begin{table*}[t!]
\centering
\begin{tabular}{|c|c|c|c|c|c|}
\hline
{ } & {Low Network Congestion} & {Medium Network Congestion} & {High Network Congestion} \\
\hline
{AVC (CRF=30)} & {100 \%} & {100 \%} & {100 \%} \\
\hline
{AVC (CRF=23)} & {100 \%} & {100 \%} & {0 \%} \\
\hline
{AVC (CRF=18)} & {100 \%} & {0 \%} & {0 \%} \\
\hline
{PNC} & {100 \%} & {100 \%} & {100 \%} \\
\hline
{FrameCorr} & {100 \%} & {100 \%} & {100 \%} \\
\hline
\end{tabular}
\caption{The video `Swing Bench 1' was randomly selected from our dataset to represent the percentage of video successfully transmitted over the network when a deadline of 300 ms is enforced for the entire video (for AVC) or 6 ms per frame (for PNC and FrameCorr), given that 'Swing Bench 1' contains approximately 59 frames, 300 ms / 59 $\approx$ 5 ms.}
\label{tab:percentage_of_vids_sent}
\end{table*}

\section{Discussion}
\textbf{No loss of information.} Our results suggest that if we can assume no loss of information, traditional video compression methods such as AVC demonstrate robust performance in reconstructing compressed data. However, training deep learning models poses its own set of hurdles. Firstly, generating a representative dataset may be impractical due to the diverse and intricate nature of real-world data sources. Secondly, model training necessitates fine-tuning hyperparameters and substantial computational resources.

If the application can accommodate data delivery delays, conventional ABR algorithms can dynamically adjust the bitrate in low-bandwidth scenarios to facilitate network data transmission (simply switch to the lower bitrate-encoded video for transfer). Consequently, we contend that in most situations, adopting a state-of-the-art traditional video compression algorithm remains the preferable approach.

\textbf{Information Loss:} Reconstructing frames with incomplete data poses a significant challenge for many video compression algorithms. Traditional methods like AVC require complete encoded information, and any partial loss can result in data corruption. In contrast, deep learning models can handle missing information through zero-padding, pixel prediction, etc. albeit with compromised reconstruction performance. Surprisingly, our study reveals that \project ~underperforms compared to the state-of-the-art method, PNC. Several factors may account for this discrepancy:

\begin{itemize}
\item \textit{Training Discrepancy:} We train \project ~to predict encoded information for a frame based on other encoded data of previous $K$ frames. However, it's possible that the encoded information space differs significantly between the training, validation, and test sets. As a result, \project ~may struggle to accurately predict the encoding for videos in the test set.
    \item \textit{Model Complexity:} Our experiments indicate that setting $K=1$ yields the best MSE value. This suggests that the simple two-layer neural network architecture used in \project ~may not adequately capture the relationships among consecutive frames. Employing more sophisticated models like LSTMs could potentially improve performance in this context.
\end{itemize}

\textbf{ Potential Workflows for \project}

Despite the limitations of \project, its deep learning-based frame reconstruction approach holds potential value for specific workflows. Consider a system severely constrained in network bandwidth (e.g., a consistently offline remote network) but possessing ample power and compute resources. In this scenario, sending highly condensed sets of features combined with reference frames, coupled with \project's ~reconstruction capabilities, could maximize the amount of usable data transmitted.  This approach leverages \project's emphasis on local computation rather than network-intensive data transfer. While the high accessibility of the internet currently limits the prevalence of such use cases, there is value in exploring maximizing the potential of limited data through reconstruction, even beyond networking-oriented scenarios.

\section{Future Work}
Although \project\ did not yield the most promising results, our extensive experimentation highlighted several opportunities to investigate alternative methodologies for improved outcomes and expand our testing into more realistic scenarios.

\textbf{Integration of FrameCorr with Adaptive Bitrate Streaming.} One potential exploration is the integration of the FrameCorr paradigm with ABR-based techniques. By combining FrameCorr's capability to reconstruct frames from partially received data with ABR's dynamic bitrate adjustment, there may be improvements in video quality and resilience to network fluctuations, particularly congested network conditions. 

\textbf{Real-world Implementation on IoT Devices.} Although our current experiments were conducted on machines hosted by a remote VM cluster, there is a need to validate our findings on actual IoT devices. Implementing and testing FrameCorr on physical IoT hardware, such as Raspberry Pi devices or other low-power embedded systems, can provide better insights into the practical challenges and performance impacts in real environments.

Furthermore, switching to live data capture or other similar workflow is another area of exploration that will not only scrutinize \project's ~flexibility in live, real-time streaming scenarios but also expose the system to greater I/O latency and storage limitations. 

\section{Conclusion}
Despite the prevalent challenge in video capture and processing systems of being unable to transmit complete data due to constraints such as limited time and bandwidth, traditional and deep-learning-based approaches appear to be somewhat ineffective in addressing this issue. Through experimentation with AVC, PNC, and ultimately our extension of PNC, \project, we found that AVC is unable to cope with partially received data. Conversely, PNC and \project ~exhibit suboptimal performance.

Additionally, factors like power consumption, network variability, and hardware specifications demand highly specialized models and setups tailored to the specific use case. We hope this paper offers guidance on navigating the trade-offs for optimal model selection in other IoT/Edge-geared designs.

We are confident that $\project$ represents a notable step forward in addressing the challenge of effectively managing incomplete data. The current outcomes can be attributed to the selection of a basic model for \project, which we believe can be remedied through the adoption of a more suitable model and meticulous infrastructure.

\bibliographystyle{plain}
\bibliography{bibliography}
\end{multicols*}
\end{document}